\documentclass[journal=jctcce,manuscript=article]{achemso}
\setkeys{acs}{maxauthors=10,etalmode=truncate,chaptertitle =true,articletitle=true}
\newcommand{\onlinecite}[1]{\hspace{-1 ex} \nocite{#1}\citenum{#1}} 
\usepackage{amssymb}
\usepackage{amsmath}
\usepackage[usenames,dvipsnames]{xcolor}
\usepackage{graphicx}
\usepackage{float}
\usepackage[normalem]{ulem}
\usepackage{subcaption}
\usepackage{natbib}
\newcommand{\bra}[1]{\left\langle #1 \right|}
\newcommand{\ket}[1]{\left|#1\right\rangle}

\newcommand{\abs}[1]{\left|#1\right|}

\usepackage{xcolor}

\usepackage{multirow}
\usepackage{bm}
\usepackage{epstopdf}
\title{Orbital Optimized Density Functional Theory for Electronic Excited States}

\author{Diptarka Hait}
\email{diptarka@berkeley.edu}
\affiliation
{{Kenneth S. Pitzer Center for Theoretical Chemistry, Department of Chemistry, University of California, Berkeley, California 94720, USA}}
\alsoaffiliation{Chemical Sciences Division, Lawrence Berkeley National Laboratory, Berkeley, California 94720, USA}
\author{Martin Head-Gordon}
\email{mhg@cchem.berkeley.edu}
\affiliation
{{Kenneth S. Pitzer Center for Theoretical Chemistry, Department of Chemistry, University of California, Berkeley, California 94720, USA}}
\alsoaffiliation{Chemical Sciences Division, Lawrence Berkeley National Laboratory, Berkeley, California 94720, USA}

\begin{document}

	\maketitle

\begin{abstract}
Density functional theory (DFT) based modeling of electronic excited states is of importance for investigation of the photophysical/photochemical properties and spectroscopic characterization of large systems. The widely used linear response time-dependent DFT (TDDFT) approach is however not effective at modeling many types of excited states, including (but not limited to) charge-transfer states, doubly excited states and core-level excitations. In this perspective, we discuss state-specific orbital optimized (OO) DFT approaches as an alterative to TDDFT for electronic excited states. We motivate the use of OO-DFT methods and discuss reasons behind their relatively restricted historical usage (vs TDDFT). We subsequently highlight modern developments that address these factors and allow efficient and reliable OO-DFT computations. Several successful applications of OO-DFT for challenging electronic excitations are also presented, indicating their practical efficacy.  OO-DFT approaches are thus increasingly becoming a useful route for computing excited states of large chemical systems. We conclude by discussing the limitations and challenges still facing OO-DFT methods, as well as some potential avenues for addressing them. 
\end{abstract}

Electronic excited states play a key role in the photophysics and photochemistry of chemical systems  \cite{turro2009principles}. Characterization of such states is consequently not only of interest from a basic science perspective, but is also critical for efficient design of photovoltaic materials, photocatalysts, lighting devices etc. The development of theoretical
methods to model electronic excitations is thus of considerable importance. Progress in these directions has been somewhat slower than comparable efforts to model the ground state, but new and exciting developments in the area have come a long way in bridging the gap. 

Like in the ground state, very accurate wave function methods are extremely useful for reliably getting accurate excitation energies and properties. However, the computational expense of such methods mostly restrict their use to development of benchmark datasets\cite{loos2020quest,chrayteh2020mountaineering} (against which more approximate methods of lower computational complexity can be assessed). Kohn-Sham density functional theory (KS-DFT) offers an excellent balance between accuracy and computational cost for ground state calculations\cite{mardirossian2017thirty,goerigk2017look}, and has greatly contributed to the increasingly widespread use of quantum chemistry. KS-DFT protocols thus appear to be the natural route for computationally efficient modeling of electronic excitations. Indeed, linear-response time-dependent DFT\cite{casida1995time,dreuw2005single} (LR-TDDFT, henceforth refered to as TDDFT) is very widely employed for this purpose. TDDFT obtains excited state energies and properties via linear-response of a ground state DFT solution to time-dependent electric fields and (within the KS formalism) is mathematically quite similar to 
TD Hartree-Fock (TDHF)\cite{dirac1930note,heinrichs1968new}. 

TDDFT is formally exact\cite{runge1984density} if the exact time-dependent exchange-correlation (xc) functional is employed. In practice, approximate, time-independent ground state xc functionals are instead utilized, (the so called adiabatic local-density approximation/ALDA\cite{dreuw2005single}). This route is capable of yielding reasonable results\cite{furche2002adiabatic,jacquemin2009extensive} but has some well-known shortcomings. ALDA restricts TDDFT to single excitations alone, making it impossible to model doubly (or higher) excited states\cite{tozer2000determination,maitra2004double,levine2006conical} and excited state bond dissociations\cite{hait2019beyond}. Furthermore, ALDA prevents post-LR orbital relaxation, which is problematic for modeling excited states that have substantially different densities than the ground state. The classic example is long-range charge transfer (CT), for which TDDFT excitation energies are strongly dependent on the fraction of HF exchange present in the functional\cite{dreuw2005single}. Similar behavior is also observed for Rydberg states\cite{tozer2000determination} and core excitations\cite{besley2010time} (effectively CT out of core orbitals). 

\begin{figure}[t!]
    \centering
    \includegraphics[width=0.7\textwidth]{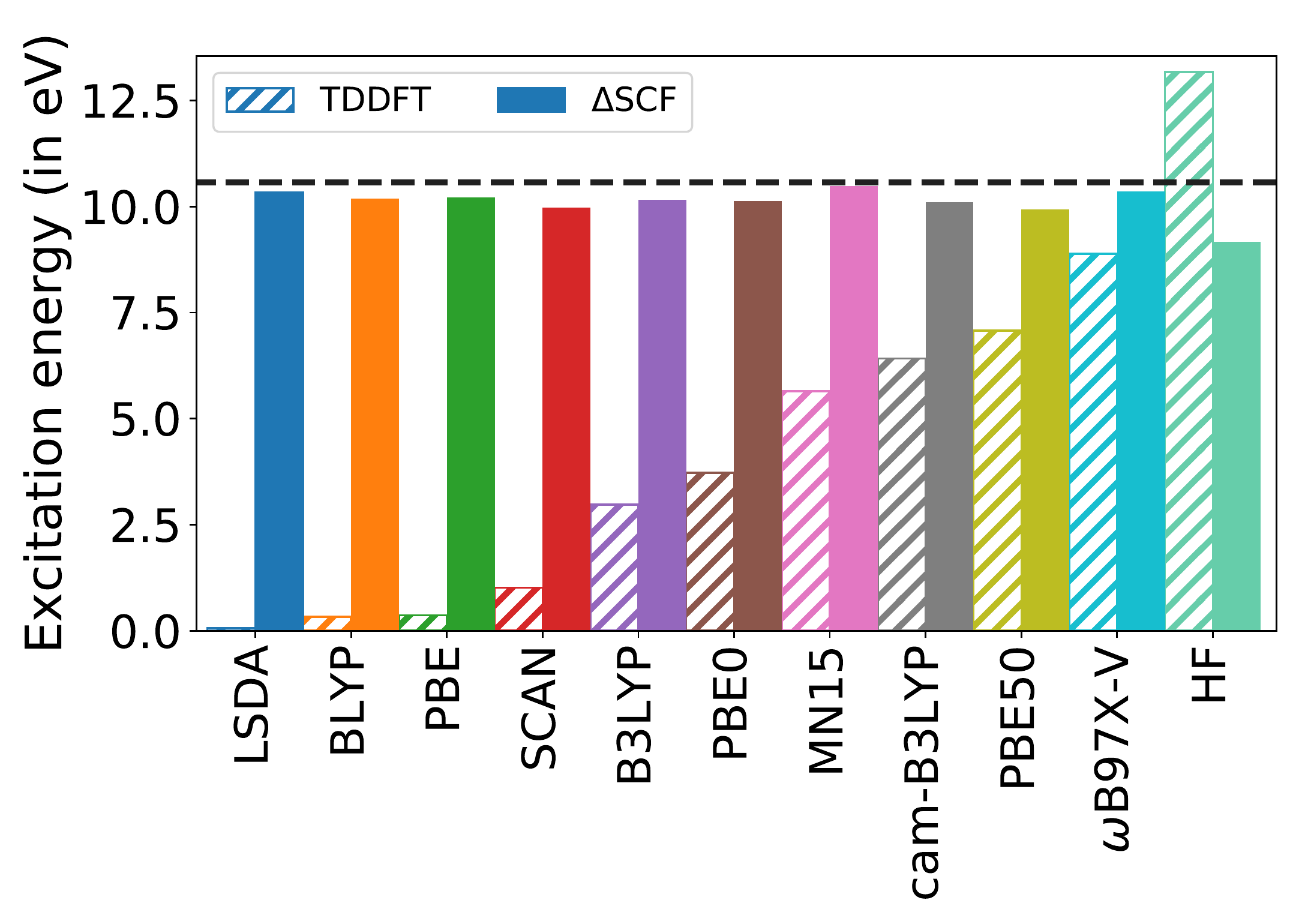}
    \caption{Comparison of TDDFT and $\Delta$SCF for the lowest CT triplet excited state of NH$_3$-----F$_2$ at 1000{\AA} separation (between the N and closest F atom), with various functionals and the def2-QZVPPD basis. The dark line represents the estimate obtained from vertical IP of NH$_3$ and vertical EA of F$_2$, evaluated via CCSD(T)\cite{raghavachari1989fifth} at the complete basis set (CBS) limit. The electrostatic interaction between unit point charges at 1000 {\AA} separation is a negligible $\sim$ 0.015 eV. }
    \label{fig:ct}
\end{figure}

The CT problem is largely a consequence of ground state delocalization error\cite{perdew1982density,hait2018delocalization}. However, the LR protocol magnifies this error to catastrophic proportions in a manner that is atypical for ground state calculations. Let us consider the lowest energy CT excitation between an electron donor $A$ and an electron acceptor $B$, at infinite separation. TDDFT predicts the excitation energy to be the difference in energy between the LUMO of $B$ and HOMO of $A$\cite{dreuw2005single}, which proves to be quite inaccurate in practice and is also quite functional sensitive (being underestimated by local functionals and overestimated by pure HF). However, we know that the true excitation energy should equal the sum of the ionization potential (IP) of $A$ and the electron affinity (EA) of $B$. Both are typically well approximated by ground state DFT\cite{mardirossian2017thirty,goerigk2017look}, indicating the LR protocol is the principal problem. We consider CT from NH$_3$ to an F$_2$ molecule 1000 {\AA} away as an example. Fig 1 shows that TDDFT predictions as a function of functional span a wide range from $\sim 0$ (TD-LSDA\footnote{We employed the SPW92\cite{Slater,PW92} functional as the local spin density approximation (LSDA) throughout.}) to $13$ eV (TDHF), as shown in Fig \ref{fig:ct}. It is worth noting that optimal tuning of density functionals\cite{kronik2012excitation} (via enforcement of Koopman's theorem\cite{szabo2012modern}) would assist in better modeling of long range CT, but would entail system-specific functional optimization and might not be as effective at intermediate separations\cite{shee2020predicting}.

Rather than using TDDFT to evaluate this excited state, we could instead consider modeling it as a supersystem consisting of NH$_3^+$ and F$_2^-$ fragments. This can be achieved by separately optimizing the orbitals of the charged fragments, constructing a guess density from these fragment densities\cite{khaliullin2007unravelling} and subsequent relaxation of the supersystem orbitals to the closest stationary point to this initial guess. The excitation energy can then be computed as the energy difference between this self-consistent field (SCF) solution and the ground state SCF solution, leading to this approach being termed as $\Delta$SCF. Fig \ref{fig:ct} shows that this orbital optimized approach yields much more reasonable results than TDDFT with the same functionals, and has much lower functional sensitivity. This indicates that state specific OO-DFT could be effective in addressing several of the challenges faced by TDDFT. Indeed, such methods precede TDDFT in the literature\cite{phillipson1958improved,hunt1969orthogonality,basch1970interpretation,bagus1975singlet,ziegler1977calculation} (dating back to at least Phllipson and Mulliken's 1958 work on the $^3\Sigma_u^+$ and $^1\Sigma_u^+$ states of H$_2$\cite{phillipson1958improved} with HF), but were not widely used after modern TDDFT implementations came into existence. In our opinion, there are three principal reasons for this:
\begin{enumerate}
    \item TDDFT is capable of simulataneously computing multiple states, without any prior knowledge about the nature of individual states. OO methods necessarily require information about the nature of target states, and are thus not very `black box'. For instance, we had to explicitly specify an initial guess of NH$_3^+$ and F$_2^-$ for the example shown in Fig \ref{fig:ct}.
    \item Excited states are typically \textit{not} minima of the energy in orbital space.  \footnote{This can be intuitively understood by noting that the Hessian of the energy vs orbital rotations for a single, spin-unrestricted determinant is $2\left(\mathbf{A}+\mathbf{B}\right)$, where $\mathbf{A},\mathbf{B}$ are the well known TDDFT matrices\cite{dreuw2005single}. Considering only mean-field one body terms, we find that the Hessian is diagonal, with the eigenvalues being differences in energy between unoccupied and occupied orbitals. Normally, there exists at least one excited state occupied orbital with higher energy than an unoccupied one (non-aufbau filling), leading to at least one negative eigenvalue for the Hessian, within this approximation. } Optimization of excited state specific orbitals is thus challenging, as it is easy to `slip' into a nearby local minimum instead of the desired state, which is often described as \textit{variational collapse}. As an example, the lowest triplet state of the NH$_3$---F$_2$ supersystem is a local excitation on F$_2$ ($\sim$ 3 eV excitation energy), and routine SCF cycles would typically land on this state instead of the desired CT state with $\sim$ 10 eV excitation energy. 
    \item Excited states are often intrinsically multireference. 
    KS-DFT is a single determinant theory by construction, and it is not straightforward to account for multiconfigurational effects. This is generally not an issue for TDDFT out of closed-shell determinants, as the LR protocol gives a formally appropriate route for `coupling' excited configurations, in a manner analogous to configuration interaction singles (CIS). \footnote{It is nonetheless worth noting that the lack of double excitations prevents open-shell TDDFT from fully addressing multiconfigurational effects, as no rigorous analogues to XCIS\cite{maurice1996nature} exist.}
\end{enumerate}

The first challenge is unavoidable in some regards, as some information about the target state is needed to provide an initial guess for orbital optimization. Specifically, it is extremely useful to know the potential electronic configurations the target state could have, out of the combinatorially scaling possibilities in the full Hilbert space.  An initial TDDFT calculation could in fact be useful in identifying the electronic configuration for singly excited states, that could be subsequently used to initiate OO calculations. Doubly (or higher) excited states are more challenging as they are inaccessible in TDDFT, and more rigorous multireference wavefunction methods capable of identifying important configurations are unlikely to be affordable for sizeable systems. 
A combination of chemical intuition and orbital energies (perhaps augmented by machine learning) could be effective in generating potential guess configurations for such excited states, if the precise nature of the desired states is not known in advance. We will consequently focus on the variational collapse and multiconfigurational problems for the rest of this work. 

The variational collapse problem has long been a major barrier to widespread use of OO methods\cite{tozer2000determination,cheng2008rydberg} beyond the lowest energy state within each spin manifold (such as the lowest energy triplet). It is possible to avert this for certain problems via application of constraints, such as the lowest energy long-range CT excitation shown in Fig \ref{fig:ct}, where individual charges on fragments are unambiguous and can be constrained\cite{kaduk2012constrained}. Similarly, a specific non-aufbau configuration can be enforced over all SCF cycles when the orbital energy ordering remains unchanged (such as insisting upon a 2s hole for an Ar ion\cite{bagus1965self}). 
However, not all problems are as clear cut---necessitating development of more general solutions. One of the most widely used approaches is the maximum overlap method (MOM)\cite{gilbert2008self}, which is used in conjunction with Fock matrix ($\mathbf{F}$) diagonalization based methods like DIIS\cite{pulay1980convergence}. Normally, the selection of occupied orbitals after each SCF cycle is done on the basis of energy, such that the lowest energy levels are filled first (aufbau principle).  
MOM instead selects occupied orbitals via maximizing the overlap between the newly constructed determinant with the determinant from the previous iteration, thereby preventing any dramatic change in the density and permitting smooth relaxation of an initial non-aufbau configuration to an optimized extrema. 

Although effective in many cases, MOM cannot always prevent variational collapse as orbitals can continuously change character back to the ground state over multiple steps\cite{mewes2014molecular,barca2018simple}. This led to the development of the Initial MOM (IMOM) method\cite{barca2018simple}, where the overlap was now maximized relative to the initial set of orbitals to prevent continuous drift down to the ground state. However, dramatic changes in electronic configuration is possible in IMOM when multiple selections can lead to similar overlaps, leading to potential oscillatory behavior and convergence failure\cite{hait2020excited,carter2020state}. An alternative route to avoiding variational collapse within the repeated $\mathbf{F}$ diagonalization scheme is through the use of level shifts\cite{saunders1973level}(as most recently exemplified by the STEP approach\cite{carter2020state}). This entails shifting the unoccupied orbitals up in energy prior to $\mathbf{F}$ diagonalization, which both permits aufbau filling of orbitals (as the undesired levels are pushed up and thus unlikely to be filled) and decelerates occupied-virtual mixing, permitting slow but steady convergence. 

Repeated $\mathbf{F}$ diagonalization based approaches however do not guarantee convergence, as is often painfully evident for nontrivial ground state computations. The most robust solvers for these problems are (quasi-) Newton schemes\cite{bacskay1981quadratically} like geometric direct minimization (GDM) \cite{van2002geometric}, that explicitly attempt to minimize the energy and thus guarantee descent, step by step. Such minimizers however are definitionally unsuited for saddle point convergence, indicating that it might be useful to transform the energy extremization problem as a minimization scheme for some other function. A natural choice in this regards is the variance ($\bm{\sigma}^2$) of the Hamiltonian $\mathbf{H}$ (i.e. $\bm{\sigma}^2=\mathbf{H}^2-\langle \mathbf{H}\rangle^2$), as every energy eigenstate corresponds to a global minimum of $\langle \bm{\sigma}^2\rangle$\cite{weinstein1934modified,umrigar1988optimized}. Recent work by the Neuscamman\cite{zhao2016efficient,shea2018communication} and Van Voorhis\cite{ye2017sigma,ye2019half} groups have examined $\mathbf{H}^2$ based approaches, obtaining very promising results. However, $\mathbf{H}^2$ is a rather difficult quantity to work with, as it contains four particle operators. Furthermore, there is no analogue of $\mathbf{H}^2$ in DFT (i.e. no functionals for $\langle\mathbf{H}^2\rangle$). 

We had consequently examined an alternative route for excited state orbital optimization that focuses on the square of the gradient of the energy $E$ vs orbital degrees of freedom $\vec{\theta}$. By defining the positive semidefinite quantity $\Delta=\abs{\nabla_{\vec{\theta}}E}^2$, we can see that all stationary points of $E$ correspond to global minima of $\Delta$ (and vice-versa). Minimization of $\Delta$ from an initial guess configuration should thus lead to the closest stationary point, as long as the gradient descent steps are sufficiently small. Furthermore, use of a general Lagrangian $\mathcal{L}$ instead of $E$ would readily permit use of this square gradient minimization (SGM) approach to \textit{any} quantum chemistry method like second order M{\o}ller-Plesset perturbation theory (MP2) or coupled cluster doubles (CCD), and not just DFT. 

The details of SGM are provided in Ref \citenum{hait2020excited}, and we only touch upon a few aspects here. SGM requires the orbital gradient $\nabla_{\vec{\theta}}\Delta$, which can be computed analytically (when analytic orbital energy hessians $\nabla^2_{\vec{\theta}}E$ are available) or via finite differences. Both approaches have similar cost (twice the evaluation of a single $\nabla_{\vec{\theta}}E$ gradient for the analytic computation, thrice for two-point centered finite-differences), although the latter is certainly easier to implement for an arbitrary quantum chemistry ansatz. This ensures that SGM preserves the scaling of ground state orbital optimization, and thus the largest problems tractable with modern day DFT should be accessible with SGM as well. A reasonable diagonal preconditioner can also be obtained by following the spirit of GDM and only preserving $\mathbf{F}$ matrix terms, which works out to be twice the square of the GDM preconditioner. Furthermore, SGM for a single determinant can be viewed as minimization of the variance of $\mathbf{F}$, permitting an interpretation of $\Delta$ as a mean-field analogue of $\bm{\sigma}^2$. $\Delta$ is also a special limit of the recently proposed generalized variational principle (GVP\cite{shea2020generalized}) for excited state OO. The GVP however contains explicit energy targeting terms (among other things) for `locking' onto desired states. SGM is thus simpler but more guess sensitive, as it only employs $\Delta$. 

$\Delta$ minimization however has two apparent shortcomings. It is less well conditioned than $E$ minimization, as it effectively squares the condition number. The preconditioner assists in partially mitigating this problem, but the convergence can still require perceptibly more iterations than MOM/IMOM (even without accounting for the higher cost per iteration). $\mathbf{F}$ diagonalization based methods could thus be optimal for initial explorations, with SGM serving as a robust alternative in the face of challenging behavior.
SGM could also get `stuck' in local minima of $\Delta$ that are not stationary points in $E$ (i.e. $\Delta \ne 0$). The analogous behavior for potential energy surfaces (vs nuclear coordinates) is well-known\cite{doye2002saddle}, but we have found it to be extremely rare for orbital optimization. These unphysical minima can be furthermore averted in practice by converging onto the right state with a different (ideally, cheaper and simpler) functional, and using the resulting orbitals as a better guess for $\Delta$ minimization with the chosen method. Overall, we note that neither of these issues have prevented us from optimizing any desired state, and seem unlikely to pose significant obstacles for practical use. 

In addition to the methods discussed above, several other recent works have proposed various routes to avoid variational collapse\cite{cullen2011formulation,evangelista2013orthogonality, ramos2018low,burton2020energy,levi2020variational,roychoudhury2020neutral}. It would be difficult to elaborate upon all of these methods in the present work, but we encourage the interested reader to investigate the literature in this area.


Having examined several potential solutions to the variational collapse problem, we next consider the problem of obtaining intrinsically multireference excited states with KS-DFT. It must be noted that not all excited states necessarily require multiple determinants for a proper representation. States in which all unpaired electrons have the same spin can in fact be represented by single determinants, with well known examples being low lying triplet states of closed shell species (like the 1s$^1$2p$_z^1$ triplet state of He), certain double excitations\cite{loos2019reference} (such as n$^2\to\left(\pi^*\right)^2$ in HCHO) and single excitations to/from singly occupied levels in open-shell systems\cite{hait2020accurate}. DFT optimization of a single Slater determinant is thus sufficient for such states, and the resulting protocol is termed as $\Delta$SCF. $\Delta$SCF is also used to describe ionization from orbitals other than the HOMO, as they lead to formation of excited states of the cation\cite{bagus1965self,carter2020state}. 

Singlet single excitations out of closed-shell molecules are however not representable by a single determinant. Mathematically, singly excited 
singlet ($\ket{\Psi_S}$) and triplet ($\ket{\Psi_T}$) states corresponding to an excitation from orbital $i$ to $a$ from a closed-shell determinant $\ket{\Phi}$ are given by:
\begin{align}
    \ket{\Psi_S}&=\dfrac{1}{\sqrt{2}}\left(a_a^\dagger a_i+a_{\bar{a}}^\dagger a_{\bar{i}}\right)\ket{\Phi}=\dfrac{1}{\sqrt{2}}\left(\ket{\Phi^a_i}+\ket{\Phi^{\bar{a}}_{\bar{i}}}\right)\\
    \ket{\Psi_T}&=\dfrac{1}{\sqrt{2}}\left(a_a^\dagger a_i-a_{\bar{a}}^\dagger a_{\bar{i}}\right)\ket{\Phi}=\dfrac{1}{\sqrt{2}}\left(\ket{\Phi^a_i}-\ket{\Phi^{\bar{a}}_{\bar{i}}}\right)
\end{align}
within the $M_s=0$ manifold ($a^\dagger$/$a$ are second quantization creation/annihilation operators\cite{szabo2012modern}). Equal contributions from both $\ket{\Phi^a_i}$ and $\ket{\Phi^{\bar{a}}_{\bar{i}}}$ is a consequence of both the up and down spins being equally likely to be excited. 
The broken symmetry $\ket{\Phi^a_i}$ determinant (when formed from spin-restricted orbitals) is thus an equal mixture of $\ket{\Psi_S}$ and  $\ket{\Psi_T}$, indicating that $\Delta$SCF is incapable of directly yielding spin-pure results. However, the triplet energy $E_T$ can be accessed from the $M_s=\pm 1$ manifold, as the resulting configuration is well represented by a single determinant ($a_a^\dagger a_{\bar{i}}\ket{\Phi}=\ket{\Phi^a_{\bar{i}}}$ for the $M_s=1$ case). The singlet energy $E_S$ can then be obtained from approximate spin-projection\cite{yamaguchi1988spin} (AP):
\begin{align}
    \ket{\Phi^a_i}&=c_1 \ket{\Psi_S}+c_2\ket{\Psi_T},  \abs{c_1}^2+\abs{c_2}^2=1\\
      \therefore 
   \langle S^2\rangle_{mixed}&= \bra{\Phi^a_i}\mathbf{S^2}\ket{\Phi^a_i}=2\abs{c_2}^2\\
   E_{mixed}&
   =\abs{c_1}^2E_S+\abs{c_2}^2E_T\\
   \implies E_S &=\dfrac{E_{mixed}-\abs{c_2^2}E_{T}}{\abs{c_1^2}}=\dfrac{2E_{mixed}-\langle S^2\rangle_{mixed}E_{T}}{2-\langle S^2\rangle_{mixed}}\label{dscf}
\end{align}
$\langle S^2\rangle_{mixed}=1$ if restricted open-shell (RO) orbitals are used, yielding:
\begin{align}
    E_S=2E_{mixed}-E_{T} \label{ROKS}
\end{align}
This corresponds to $\ket{\Phi^a_i}$ being precisely halfway between $ \ket{\Psi_S}$ and $ \ket{\Psi_T}$ in energy.  In practice, use of spin-unrestricted (U) orbitals leads to additional spin contamination in $\ket{\Phi^a_i}$, causing $\langle S^2\rangle_{mixed}$ to be slightly larger than 1. However, there are also some species where a low-lying, closed-shell configuration can potentially mix with the singly excited state when U orbitals are used, leading to $\langle S^2\rangle_{mixed} < 1$.  $E_{mixed}$ and $E_T$ are typically separately optimized, followed by AP using either Eqn \ref{dscf} or \ref{ROKS}. Eqn \ref{ROKS} in particular has been long been used in the literature\cite{ziegler1977calculation,bagus1975singlet,kowalczyk2011assessment}, as most low lying singlet excited state of closed-shell species are of pure open-shell character. For such cases, $E_S>E_{mixed}>E_T$ is the usual ordering
due to lack of an exchange stabilization term in the mixed configuration (vs the triplet). 

It is however worth noting that some authors elect against carrying out the AP protocol, effectively arguing that $E_{mixed}\approx E_S$\cite{gilbert2008self,besley2009self,barca2017excitation,barca2018simple}. This is fairly reasonable when the unpaired electrons in orbitals $i$ and $a$ interact very weakly (very long range CT being an obvious example), but could lead to systematic underestimation of $E_S$ in general. However, this is at times compensated by errors in the xc functional, permitting fortuitously reasonable results. We feel that the AP protocol should be carried out for $\Delta$SCF calculations for singlet excitations with two unpaired electrons, in order to have a reasonably spin-pure result. 

The AP scheme described above however results in two separate orbital optimizations for $E_{mixed}$ and $E_T$. For purely open-shell singlet states, it is instead possible to directly optimize Eqn. \ref{ROKS} for a single set of RO orbitals\cite{frank1998molecular,filatov1999spin,kowalczyk2013excitation}. This has been described as restricted open-shell Kohn-Sham (ROKS) in the literature\cite{kowalczyk2013excitation}, although it is distinct from normal RO KS-DFT calculations of open-shell systems (where all unpaired spins point the same way). The HF variant of ROKS has a long history of use\cite{phillipson1958improved,hunt1969orthogonality,basch1970interpretation} (often being described as "Open-Shell SCF" in older literature), as it represents a minimal, single configuration state function (CSF) approximation to open-shell singlets within multiconfigurational SCF (MCSCF) theory. This single CSF property renders ROKS incapable of accounting for any closed-shell character in the target states, but this in turn prevents \textit{complete} variational collapse down to the lowest energy, closed-shell (S$_0$) configuration. Variational collapse in ROKS is thus normally to the lowest excited singlet (S$_1$) state. Analytic nuclear gradients for ROKS are also known\cite{kowalczyk2013excitation}, permitting excited state geometry optimizations and (finite-difference) frequency computations.

Both AP-$\Delta$SCF and ROKS are however only applicable to singlet states with one broken electron pair. The utility of AP is limited for cases with a larger number of unpaired electrons, as each mixed determinant is subsequently a combination of more than two states\cite{ziegler1977calculation}, with several having the same $\langle S^2\rangle$. Recoupling of mixed determinants to yield spin-pure energies can instead be done in a manner described in Ref \citenum{hait2020accurate} (which is rigorous for restricted open-shell HF orbitals).
We present the case with four unpaired electrons as a representative example. Let these unpaired electrons be in orbitals labeled 1-4. There are 16 possible RO determinants, but only 8 unique HF energies:
\begin{enumerate}
    \item $\ket{HS}=\ket{\uparrow\uparrow\uparrow\uparrow}$ with energy $E_{HS}$.
    \item $\ket{M_1}=\ket{\downarrow\uparrow\uparrow\uparrow}$ $E_{M_1}=E_{HS}+K_{12}+K_{13}+K_{14}$
    \item $\ket{M_2}=\ket{\uparrow\downarrow\uparrow\uparrow}$, $E_{M_2}=E_{HS}+K_{12}+K_{23}+K_{24}$.
    \item $\ket{M_3}=\ket{\uparrow\uparrow\downarrow\uparrow}$, $E_{M_3}=E_{HS}+K_{13}+K_{23}+K_{34}$.
    \item $\ket{M_4}=\ket{\uparrow\uparrow\uparrow\downarrow}$, $E_{M_4}=E_{HS}+K_{14}+K_{24}+K_{34}$.
    \item $\ket{M_5}=\ket{\downarrow\downarrow\uparrow\uparrow}$, $E_{M_5}=E_{HS}+K_{13}+K_{23}+K_{14}+K_{24}$.
    \item $\ket{M_6}=\ket{\downarrow\uparrow\downarrow\uparrow}$, $E_{M_6}=E_{HS}+K_{12}+K_{23}+K_{14}+K_{34}$.
    \item $\ket{M_7}=\ket{\downarrow\uparrow\uparrow\downarrow}$, $E_{M_7}=E_{HS}+K_{12}+K_{13}+K_{24}+K_{34}$.
\end{enumerate}
where $K_{pq}$ is the exchange interaction between electrons in orbitals $p$ and $q$. The $M_s=1$ subspace consists of $\ket{M_{1,2,3,4}}$, which combine to yield the HS quintet state, and three triplet states. The $M_s=0$ space is similarly composed of $\ket{M_{5,6,7}}$ and the determinants that arise from inverting their spins, which combine to yield two singlet states in addition to the aforementioned quintet and three triplets. 

The expressions for $E_{M_i}$ show that it is possible to solve for $K_{ij}$ from the single determinant energies $E_{HS}$ and $E_{M_i}$ (albeit via an overdetermined linear system with 7 equations and 6 unknowns). Furthermore, trivially non-zero off-diagonal $\mathbf{H}$ elements within this subspace of states are various $-K_{ij}$ (from Slater-Condon rules\cite{szabo2012modern}). In other words, we have $\bra{M_1}\mathbf{H}\ket{M_2}=-K_{12}$, $\bra{M_5}\mathbf{H}\ket{M_6}=-K_{23}$ etc. It is thus possible to obtain spin-pure energies by diagonalizing $\mathbf{H}$ within the $M_s=1$ and $M_s=0$ subspaces, with no more information than the single-determinant energies $E_{HS}$ and $E_{M_i}$. We note that Davidson had made a similar observation earlier\cite{davidson1973spin}. 

This CI based recoupling protocol using HF energies can be `extended' to DFT via use of KS energies instead of HF. The $K_{ij}$ cease to be interpretable as exchange interactions but become effective spin-spin coupling constants instead. Furthermore, the overdetermined linear system no longer has an exact solution but can be `solved' via least-squares. This approach yields Eqn \ref{ROKS} when applied to the case of two unpaired electrons and thus serves as a generalization of  Eqn \ref{ROKS} to more complex cases. However, simple analytic expressions for spin-pure energies are no longer possible, and ROKS style optimization of a single set of orbitals is thus much more challenging. It is therefore much easier to individually optimize $\ket{HS}$ and $\ket{M_{i}}$ directly for carrying out the entire protocol (analogous to AP-$\Delta$SCF), despite the derivation utilizing a single set of RO orbitals. It is thus reasonable to view AP-$\Delta$SCF and this general recoupling as functionally `projection after variation' (while ROKS is `variation after projection'). 

The four electron case is of special importance as it is relevant for states resulting from transient absorption spectroscopic experiments (i.e. single excitations from singly excited states). However, the general case of $N$ unpaired electrons would have $2^N$ possible determinants (but only $2^{N-1}$ energies due to spin-inversion symmetry) and $\dfrac{N(N-1)}{2}$ pairwise spin-spin coupling constants, indicating a greatly overdetermined system of equations. For simplicity, it would be sufficient to only consider the $M_s=\dfrac{N}{2}$ (1 determinant), $M_s=\dfrac{N}{2}-1$ ($N$ determinants) and $M_s=\dfrac{N}{2}-2$ ($\dfrac{N(N-1)}{2}$ determinants) subspaces for finding the coupling constants (via solving an overdetermined system of equations) in the large $N$ ($>4$) limit, followed by diagonalization in the relevant subspace of interest. The least squares error for the overdetermined linear system would serve as an internal metric for reliability. 

There have been some other alternative schemes proposed in literature for recoupling mixed determinants, such as the DFT generalization of excited state mean-field (ESMF)\cite{zhao2019density} and the Becke exciton model\cite{becke2018singlet}. These approaches have been however only utilized for the case of singlet excited states with two unpaired spins, with more challenging cases being unexplored to the best of our knowledge. More routes are possible from a wave function perspective, including exact spin-projection\cite{jimenez2012projected}, half-projection\cite{ye2019half,cox1976half} and eXtended CIS (XCIS)\cite{maurice1996nature}, which are nonetheless not readily generalizable to DFT. 

We have considered a number of theoretical/computational issues about OO-DFT approaches to excited states, and it is natural to wonder how well these models fare in practice. We will consider a number of problems that have been recently been studied by OO-DFT (both by us and other groups) to highlight the efficacy of these methods. 

\begin{figure}[htb!]
    \centering
    \includegraphics[width=\textwidth]{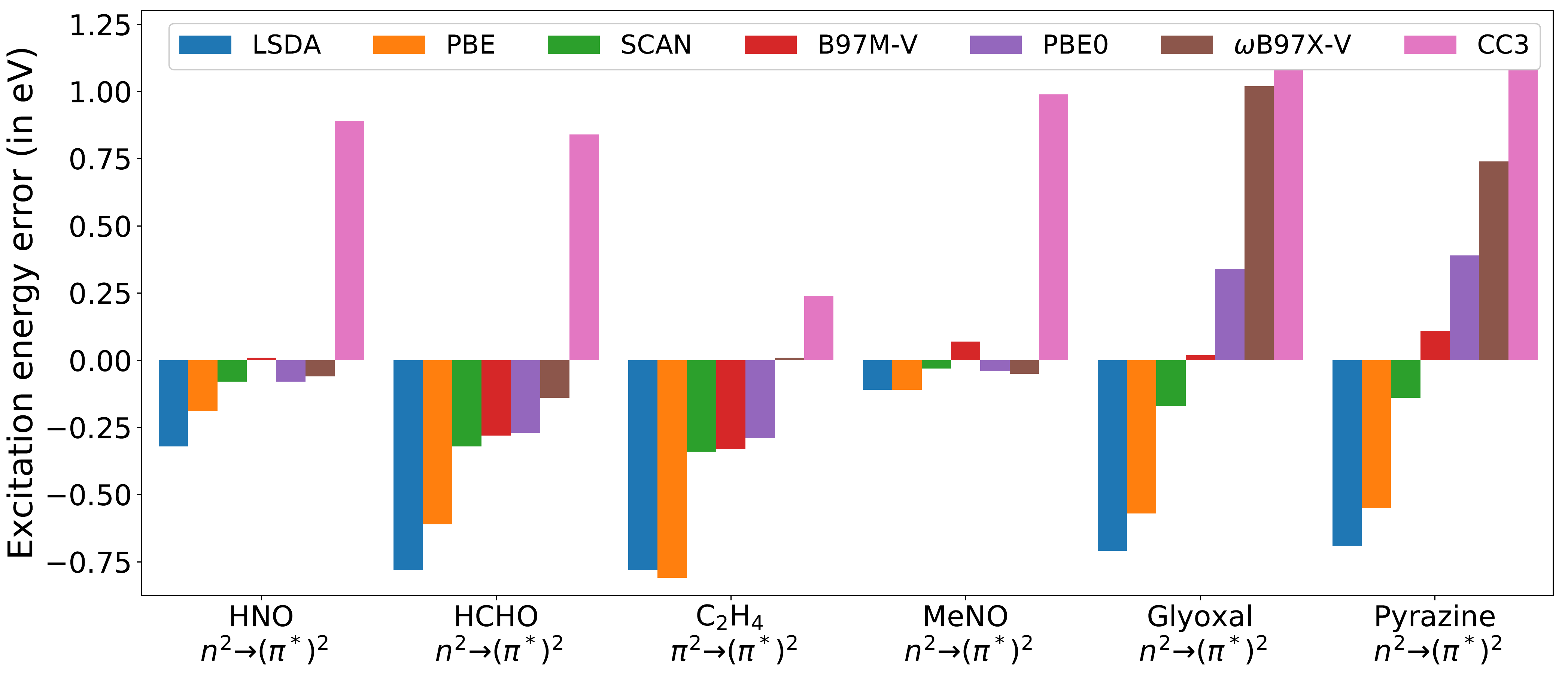}
    \caption{Errors predicted by $\Delta$SCF/aug-cc-pVTZ for a few doubly excited states, relative to reference values from Ref \citenum{loos2019reference}. CC3 results from Ref \citenum{loos2019reference} are also supplied for context. }
    \label{fig:double}
\end{figure}

Double excitations are entirely inaccessible in TDDFT, and are also poorly described by more computationally demanding CC approximations like EOM-CCSD. This has been viewed mostly as a consequence of significant multireference character of such states. However, Ref \citenum{barca2018simple} had proposed that the so-called multireference character of many doubly excited states is a spurious consequence of using ground state orbitals. Ref \citenum{loos2019reference} subsequently presented a high level wave function theory benchmark for some doubly excited states, enabling assessment of performance by OO-DFT. We examined the performance of $\Delta$SCF for closed-shell, single-determinant doubly excited states \cite{hait2020excited}, which revealed that OO-DFT was readily capable of achieving fairly accurate results (as shown by some examples in Fig \ref{fig:double}). The best accuracy is provided by the modern meta-GGAs (mGGAs) SCAN\cite{SCAN} and B97M-V\cite{b97mv} and the PBE0\cite{pbe0} hybrid functional. Even the elementary LSDA\cite{Slater,PW92} functional was able to surpass the accuracy of the $O(N^7)$ scaling CC3 method. Ref \citenum{carter2020state} subsequently reached similar conclusions as well. 

\begin{figure}[htb!]
\begin{minipage}{0.48\textwidth}
    \centering
    \includegraphics[width=\linewidth]{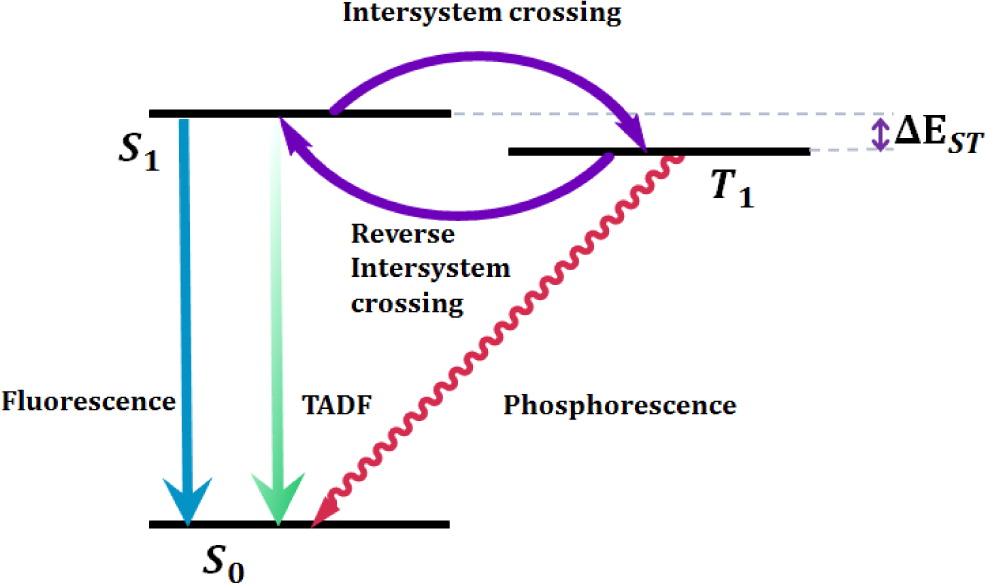}
    \vspace{15pt}
    \subcaption{Simplified Jablonski diagram from TADF (taken from Ref \citenum{hait2016prediction}). Small $\Delta$E$_{ST}$ permits reverse intersystem crossing from T$_1\to$ S$_1$, followed by fluorescence back to S$_0$.}
    \label{fig:tadfcartoon}
\end{minipage}
\begin{minipage}{0.48\textwidth}
    \centering
    \includegraphics[width=\linewidth]{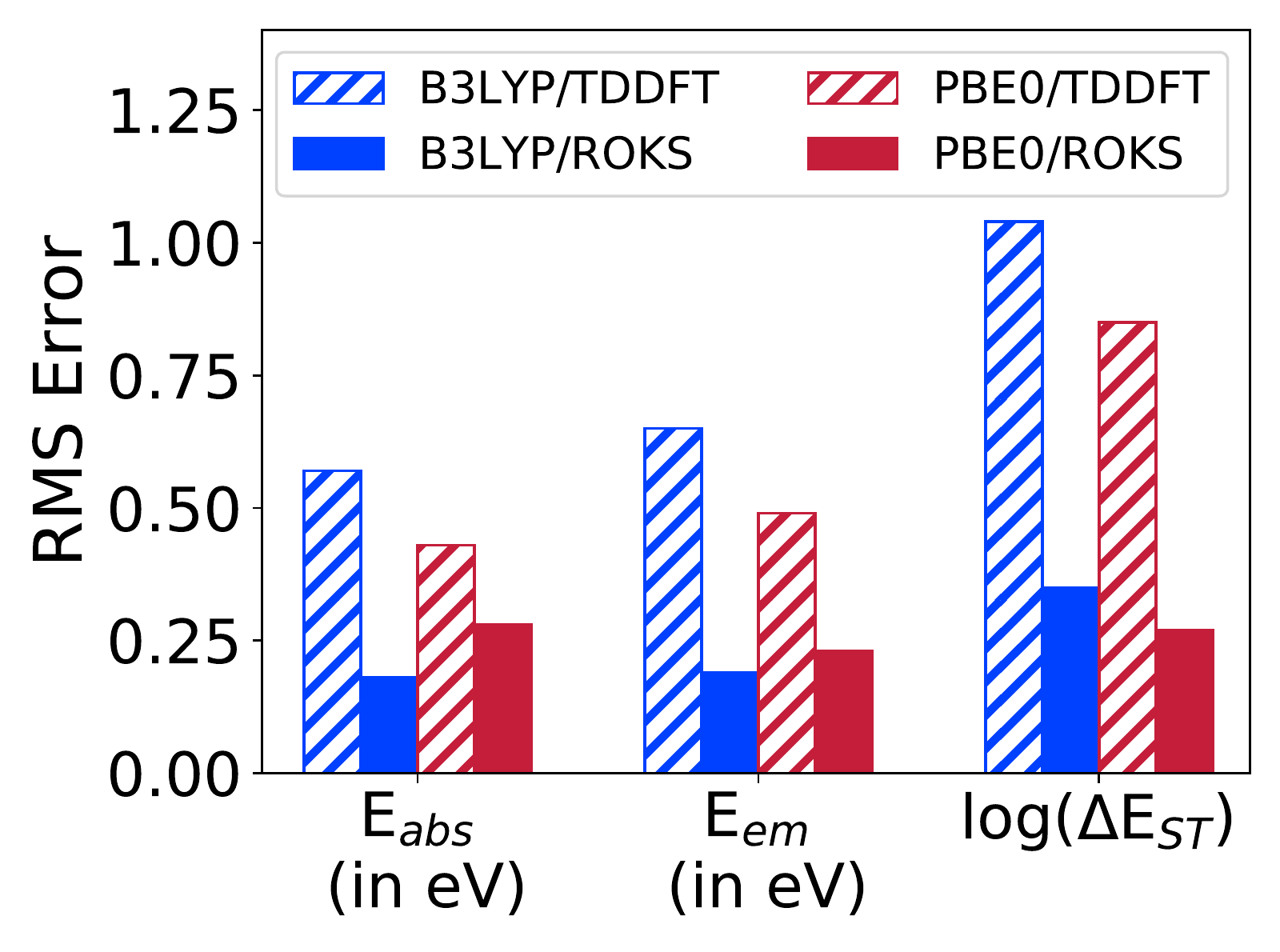}
    \subcaption{RMS errors vs experiment for the dataset in Ref \citenum{hait2016prediction}. Quantities considered are the S$_1$ absorption energy (E$_{abs}$), S$_1$ emission energy (E$_{em}$) and the $\log$ of the singlet-triplet gap $\Delta$E$_{ST}$.}
    \label{fig:tadferrors}
\end{minipage}
\caption{Thermally Activated Delayed Fluorescence (TADF).}
\end{figure}

We had previously considered an example of long ranged CT in Fig \ref{fig:ct}. Less extreme examples include molecules that exhibit thermally activated delayed fluorescence (TADF\cite{valeur2003molecular}). Species exhibiting TADF have very small singlet-triplet gaps ($\Delta$E$_{ST}$), permitting reverse intersystem crossing from triplet to singlet, followed by fluorescence back to the ground state (as shown in Fig. \ref{fig:tadfcartoon}). TADF molecules are of interest for lighting applications as it is a route to harvest energy that would otherwise be wasted in nonradiative channels by triplet excitons\cite{endo2009thermally}. Small $\Delta$E$_{ST}$
indicate weak electron-hole interaction, which is a characteristic of CT excitations, making that an important design principle. Ref \citenum{hait2016prediction} investigated the S$_1$ states of TADF molecules with ROKS. The results indicated that ROKS yielded low error (vs experimental values) for S$_1$ emission energies and especially for the small adiabatic $\Delta$E$_{ST}$ with the classic B3LYP\cite{b3lyp} and PBE0 global hybrid functionals (as shown in Fig \ref{fig:tadferrors}).  In constrast, TDDFT with these functionals fared much worse, due to systematic underestimation of CT excited state energies. Furthermore, Ref \citenum{barca2018simple} showed that $\Delta$SCF obtains the correct asymptotic behavior ($O(r^{-1})$) of CT excitation energies with distance, unlike TDDFT. ROKS should also show the correct asymptotic behavior, as it is asymptotically identical to $\Delta$SCF for long range CT. OO-DFT approaches thus appear to be quite well suited for CT problems that are difficult for TDDFT.

How about low lying singly excited states of small molecules? Ref \citenum{ye2020self} reported the performance of ROKS for a large dataset of 104 such excitations (selected from the datasets of Ref \onlinecite{loos2018mountaineering} and \onlinecite{loos2020mountaineering}), which are mostly valence or Rydberg in character. ROKS with $\omega$B97X-V\cite{wb97xv} has an RMS error (RMSE) of 0.24 eV, vs 0.13 eV from EOM-CCSD, which is very promising. ROKS with LSDA, PBE\cite{PBE} and B3LYP fares worse, with RMSE $\sim$0.40-0.55 eV, which nonetheless are pretty comparable to TDDFT errors for such systems. It is worth noting that ROKS with HF is much worse (0.81 eV RMSE), indicating that the xc contribution is crucial for such excited states. 
Of course, use of ROKS is inefficient (vs TDDFT) for valence excitations but the overall accuracy of the approach indicates that it is not just a niche method useful only when TDDFT fails. Ref \citenum{ye2020self} furthermore clearly shows a significant reduction in error on using ROKS vs $\Delta$SCF without AP, demonstrating the utility of approximate enforcement of spin-purity in practice.

\begin{figure}[htb!]
    \centering
    \includegraphics[width=0.5\textwidth]{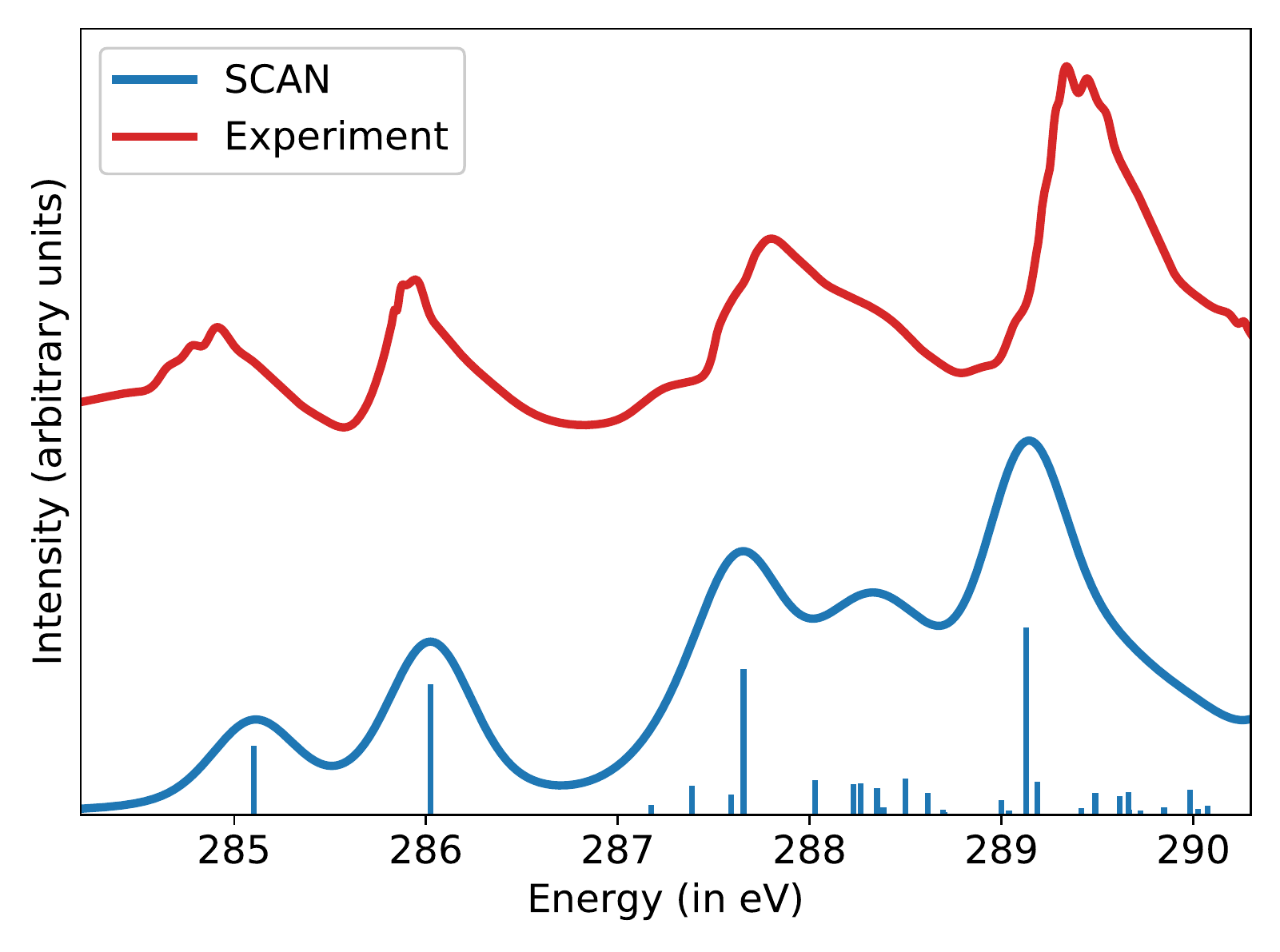}
    \caption{Comparison of computed (ROKS with SCAN/aug-cc-pCVTZ) C K-edge spectrum of thymine with experiment\cite{plekan2008theoretical}, without any empirical translation of spectra. The peak position for the highest energy peak shows the most deviation (0.25 eV, 289.40 eV vs 289.15 eV). The computed peaks were broadened by a Voigt profile with a Gaussian standard deviation of 0.2 eV and Lorentzian $\gamma$ = 0.121 eV.}
    \label{fig:thymine}
\end{figure}

A very promising area of application for OO-DFT methods is spectroscopy of core electrons. Excitations out of inner shells lead to substantial reorganization of the total electron density (the remaining core electron can have a more compact orbital due to reduced repulsion, not to mention the relaxation of the valence density in response), making them challenging for LR methods. TDDFT errors can readily span from underestimation by $\sim$10 eV\cite{besley2010time} to overestimation by $\sim$10 eV (in the pure HF limit\cite{oosterbaan2018non}), necessitating empirical translation of spectra for agreement with experiment (and motivating the development of specialized functionals optimized solely for core-level excitations\cite{besley2009time}). Even EOM-CCSD spectra often need to be shifted by a smaller amount ($\sim$ 1-2 eV) to align experimental peaks with computed ones\cite{frati2019coupled}. OO methods thus appear to be well suited here, as was noted via early applications of $\Delta$SCF using HF\cite{broer1981broken}. Ref \citenum{besley2009self} reported good results from $\Delta$SCF for core-excitations (without AP) with B3LYP. Ref \citenum{hait2020highly} subsequently reported even better results via use of ROKS, especially with the modern SCAN mGGA. In particular, SCAN obtains an RMSE of 0.2 eV vs experiment for 40 core excitations out of 1s orbitals (K-edge) of C,N,O and F in small molecules, which is quite low relative to experimental uncertainties of $\sim$ 0.1 eV. Similar accuracy is also attained for 20 excitations out of 2p orbitals (L-edge) of Si, P, S and Cl. Transition dipole moments computed from treating the KS determinants as pseudo wave functions are also quite reasonable, as shown by the good agreement between experimental and computed spectra\cite{hait2020highly}. A representative case (C K-edge of thymine) is depicted in Fig \ref{fig:thymine}.

The origin of the very good behavior of ROKS for core excitations is however closely connected to orbital optimization, as opposed to any specific choice of density functional. HF alone predicts an RMSE of $0.6$ eV vs experiment for the 40 K-edges considered in Ref \citenum{hait2020highly} (much better than mGGA functionals like B97M-V, that have RMSE $>$ 1 eV). Modern semi-empirical functional development mostly entails training and selection based on chemically relevant ground state energy differences, and there is thus little reason to believe that functionals with highly flexible forms would necessarily be successful so far from their training regime. As a corollary, the simpler semi-empirical hybrid GGA $\omega$B97X-V outperforms vastly more complex forms like the $\omega$B97M-V\cite{wB97MV} semiempirical hybrid mGGA (RMSE 0.2 eV vs 1 eV). 
Within the mGGA space, SCAN is a mostly non-empirical functional employing a large number of exact constraints, which perhaps allows it to improve over HF (via inclusion of dynamical correlation) instead of causing harm. Ultimately, the low computational cost of the SCAN mGGA permits ready applicability to large systems like porphyrin\cite{hait2020highly}, with fairly high accuracy. 
\begin{figure}[htb!]
    \centering
    \begin{minipage}{0.48\textwidth}
    \includegraphics[width=\textwidth]{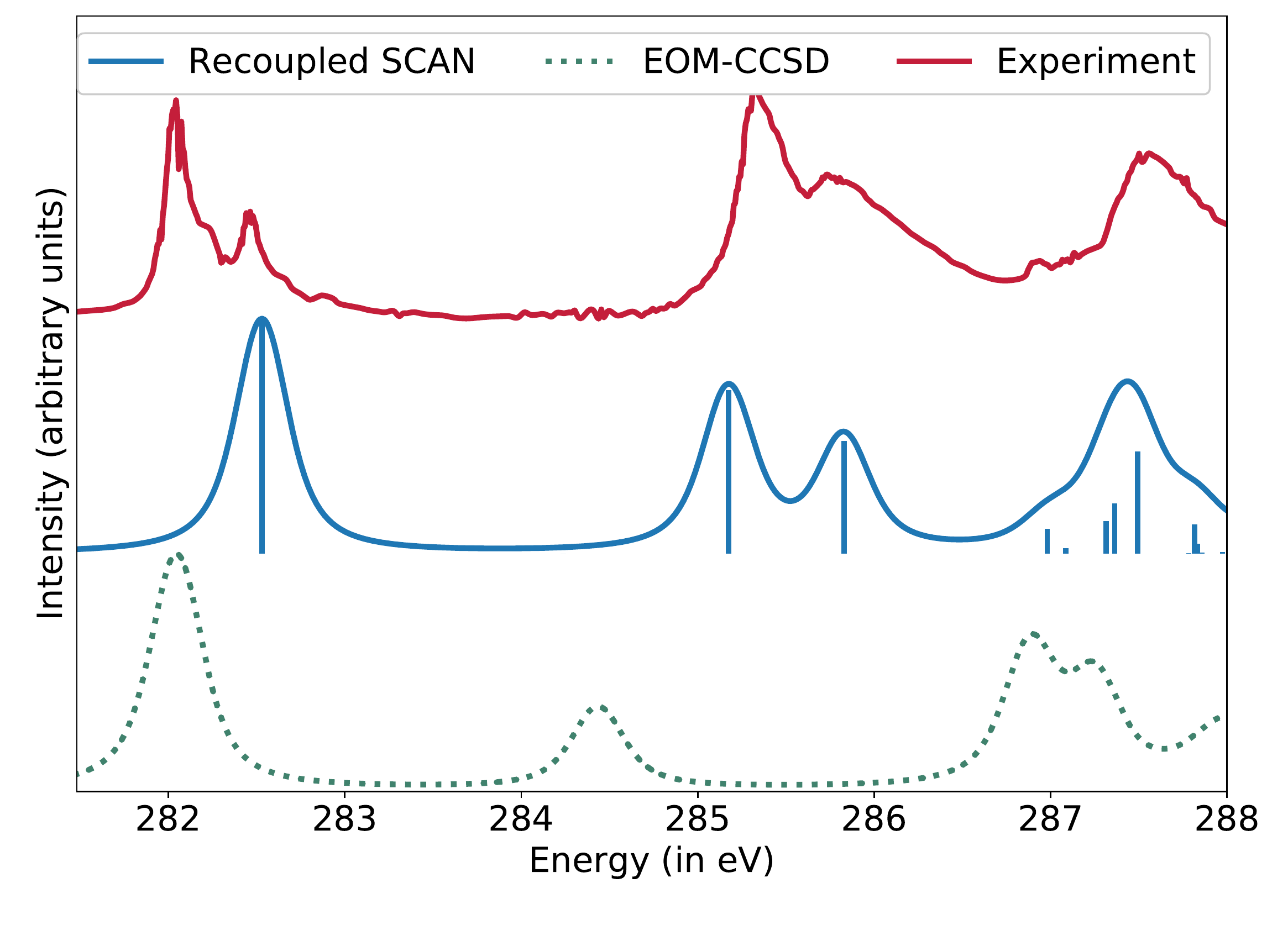}
    \subcaption{C K-edge of allyl radical.}
    \end{minipage}
    \begin{minipage}{0.48\textwidth}
    \includegraphics[width=\textwidth]{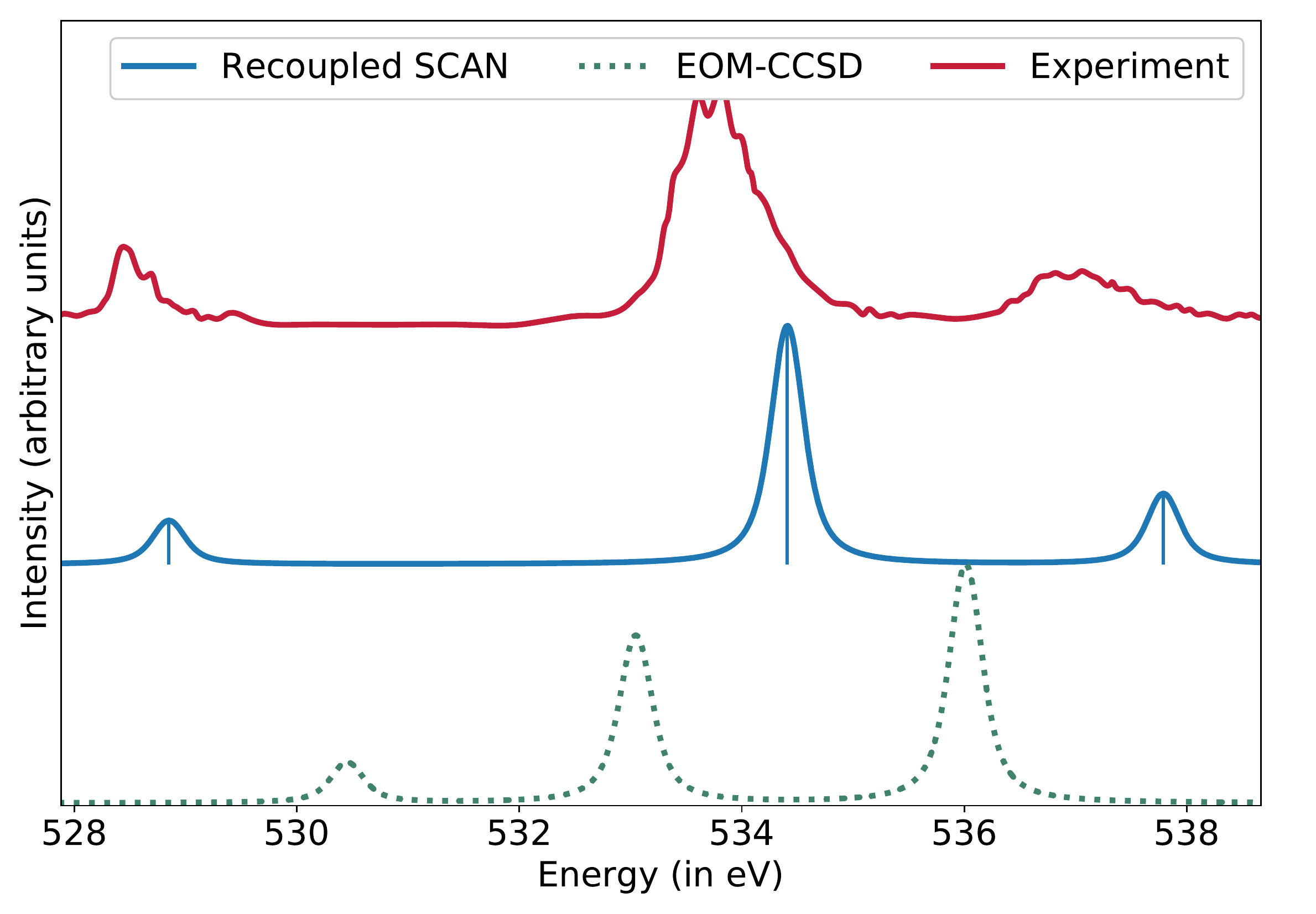}
    \subcaption{O K-edge of CO$^+$.}
    \end{minipage}
    \caption{Comparison of computed (recoupled OO-DFT with SCAN and fc-CVS-EOM-CCSD\cite{vidal2019new}, with the aug-cc-pCVTZ basis) spectra with experiment, without any empirical translation of spectra. The computed peaks were broadened by a Voigt profile with a Gaussian standard deviation of 0.2 eV and Lorentzian $\gamma$ = 0.121 eV. These results were first reported in Ref \citenum{hait2020accurate}.}
    \label{fig:osxas}
\end{figure}

Core spectra of open-shell systems also offer a chance to examine the efficacy of the presented recoupling scheme for $>2$ unpaired electrons. However, there are very few experimental spectra of unambiguously open-shell species, making it difficult to carry out comparisons. Ref \citenum{hait2020accurate} examines a few such cases for radicals (doublets with $3$ unpaired electrons), determining that recoupling appears to do no harm and in fact often leads to better agreement with experiment (over using mixed determinant energies). This makes recoupled OO-DFT very attractive for computing core-level spectra of open-shell systems, as LR methods like TDDFT/EOM-CCSD yield spin impure states when there is a net breaking of electron pairs, leading to rather suboptimal results at times. Comparison between OO-DFT and EOM-CCSD for the allyl radical and the CO$^+$ cation is supplied in Fig \ref{fig:osxas}, which indicates quite good performance by the former (albeit with perceptible room for improvement). 

In light of the aforementioned successes, it is reasonable to wonder what are the shortcomings of the OO-DFT methods. The most obvious one is the lack of a  Hohenberg-Kohn theorem for excited states\cite{gaudoin2004lack}, which indicates that there is no one-to-one mapping between the excited state density and the external potential (i.e. multiple external potentials can lead to the same excited state density). However, it has been shown\cite{perdew1985extrema} that every extremal density $\rho(\vec{r})$ of the exact ground state energy functional $E_v[\rho(\vec{r})]$ corresponds to an exact energy eigenstate (although the converse is not true, and not all excited states have densities that are stationary points of $E_v[\rho(\vec{r})]$). In fact, if $\rho_i(\vec{r})$ is an extremum of $E_v$, $E_v[\rho_i(\vec{r})]$ corresponds to the energy of the lowest energy state with density $\rho_i(\vec{r})$. This does indicate that use of OO-DFT approaches to extremize approximate ground state functionals could be effective in targeting excited states, as long as there is no state with the same density that is of lower energy. For instance, the lowest energy core-excited state should be well suited for OO-DFT, as there should not be any lower energy state that has similar density (due to the presence of the core-hole). In practice therefore, states with a well-defined, unique electronic configuration (which could aid in having an unique density), that are ``low energy" (likely characterizable by physical constraints such as a core-hole or subsystem charge) would be best approximated by OO-DFT. An alternative perspective would be to view the xc functional as merely a tool for adding dynamic correlation to a truncated wave function. Therefore, any state which can be reasonably approximated by one/few determinants that can be extremized via wave function theory, should be fairly well suited to OO-DFT as a qualitatively similar extremum is likely to be present and the xc functional only contributes a minor (but often chemically useful) correction to the energy. In fact, this can be viewed as a reason why ground state KS-DFT has been immensely successful for single-reference species (where a single determinant from HF is a good approximation). As a corollary, excited states for which a reasonable truncated wave function model cannot be developed (or optimized) are likely to be challenging for OO-DFT approaches based upon KS functionals. 

A related limitation is that approximate energy functionals are typically developed specifically for the ground state, and therefore need not be successful in modeling excited state extrema (especially ones that are very different from the ground state minima). This becomes quite apparent with more flexible functional forms like B97M-V and $\omega$B97M-V, that are very effective for ground state energies\cite{mardirossian2017thirty} but are greatly challenged by core excitations. It would therefore be interesting to see if effective density functionals explicitly designed for excited state extrema can be developed, and if they would lead to significant improvement in accuracy over existing ground state functionals like SCAN that appear to perform fairly well for excited state OO-DFT.



The multiconfigurational nature of certain excited states leads to an additional challenge for ROKS. ROKS requires that the singly occupied orbitals of the excited state be distinct from the doubly excited orbitals, and would not be able to handle states that are superpositions between different excitations. In other words, a state of the form $\dfrac{c_1}{\sqrt{2}}\left(\ket{\Phi^a_i}+\ket{\Phi^{\bar{a}}_{\bar{i}}}\right)+\dfrac{c_2}{\sqrt{2}}\left(\ket{\Phi^b_j}+\ket{\Phi^{\bar{b}}_{\bar{j}}}\right)$ would not be representable with ROKS if both $a\ne b$ and $i\ne j$ (however, OO would ensure that things are fine\cite{thouless1960stability} if either $a=b$ or $i=j$). The TDDFT solutions corresponding to these states have significant contributions from multiple natural transition orbitals (NTOs\cite{martin2003natural}), indicating ``essential configuration interaction". The presented recoupling scheme shares the same restriction in that it requires a specific set of singly occupied orbitals, and cannot be applied to excitations that involve distinct sets of singly occupied levels. 
More general ways to recouple multiple KS determinants therefore need to be explored in order to model such states. In the interim, the best option for modeling such states is to use the dominant transition (or NTO pair), in the hope that the effect of the other contributors would be folded in via the xc functional (much like ground state KS-DFT applied to multireference species). 

The OO problem with approximate functionals leads to an additional challenge in that there are often more roots than the total number of states in Hilbert space\cite{burton2020energy} (analogous to the existence of multiple local extrema for multiconfigurational SCF\cite{helgaker2014molecular}). This effect can be observed quite clearly for core electrons for symmetric molecules like CO$_2$. The canonical molecular orbitals involving the O 1s orbitals are the gerade and ungerade combinations of these atomic orbitals. O K-edge excited states thus should have either of these molecular orbitals as the core-hole. Unfortunately, delocalization error of approximate xc functionals leads to a systematic underestimation of the energy of states with delocalized core-holes\cite{hait2020highly}. Use of HF instead leads to systematic overestimation on account of overlocalization error from lack of correlation, as has been long known.\cite{broer1981broken}. However, the O 1s orbitals are essentially noninteracting and it is possible to converge orbital-optimized solutions where the core-hole is entirely localized on one O atom. The resulting solutions are symmetry broken, but yield very reasonable answers on account of being delocalization free\cite{broer1981broken,hait2020highly}. Many OO approaches involving core-excitations consequently attempt to localize the core-hole, even if alternate solutions exist. 
However, such a clear solution might not exist for many other applications, where it is not quite evident which of multiple possible OO solutions represent the best choice.  Construction of optimal guesses for initiating OO calculations is thus an important task, as it would determine the nature of the final solution. Effective partitioning of the virtual space into antibonding/Rydberg spaces (as well as proper localization of antibonding levels) could be useful in this regard, as it would permit optimization of very specific diabatic states that can be intuitively interpreted. Further work is however needed to fully characterize the dependence of the final OO solution on the nature of the initial guess. 

It is also worth noting that very little work has been done with excited state specific OO-DFT with double hybrid functionals, despite such functionals representing the most accurate ground state functionals\cite{mardirossian2018survival,santra2019minimally}. This is likely a consequence of two factors: the greater computational cost of double hybrid functionals and the remarkably slow convergence of perturbation theory for spin-contaminated references\cite{gill1988does}. However, encouraging results have been recently obtained with MP2 on top of excited state configurations optimized with HF\cite{carter2020state,ye2020self}. This suggests that better results can potentially be obtained from double hybrid functionals (when computationally affordable), especially as OO-DFT orbitals are likely to be superior to OO-HF\cite{rettig2020third}. 


In summary, the purpose of this article has been to demonstrate that state-specific orbital optimized DFT (OO-DFT) methods are remarkably accurate for a range of excited states that include low-lying valence excitations, high-lying core excitations, some doubly excited states, and low-lying charge-transfer excited states. Orbital optimization makes a very simple functional form appropriate for these applications in a way that is also physically interpretable (for instance collective expansion or contraction of orbitals when charge is moved from core to valence or donor to acceptor). OO-DFT with a single determinant ($\Delta$SCF) is appropriate for triplet excitations, some doubly excited states of closed shell molecules, and some states of open shell systems. Similarly, OO-DFT with a fixed superposition of two determinants (which is the so-called ROKS method) is appropriate for singlet excitations from closed shell ground states. We also discussed the generalization of ROKS to recouple multiple determinants for low-spin excitations in open shell systems. It is important to emphasize that new developments in state-targeted orbital optimization\cite{ye2017sigma,zhao2016efficient,hait2020excited,shea2018communication,shea2020generalized,barca2018simple,carter2020state,cullen2011formulation,evangelista2013orthogonality, ramos2018low,burton2020energy,levi2020variational,roychoudhury2020neutral}, such as our recent SGM method,\cite{hait2020excited} were essential in enabling this recent flourishing of OO-DFT. Indeed, with such algorithms to complement the seminal MOM method,\cite{gilbert2008self} the compute requirements of OO-DFT are not much worse than for ground state DFT. OO-DFT is emerging as a usable computational tool whose utility (and limitations) can now be fully revealed through large-scale applications with its availability in widely distributed software such as the latest version of Q-Chem.\cite{QCHEM4}

\section*{Acknowledgment} 
This work was supported by the Director, Office of Science, Office of Basic Energy Sciences, of the U.S. Department of Energy through the Gas Phase Chemical Physics Program, under Contract No. DE-AC02-05CH11231.  Additional support for D.H. during the preparation of this article came from the Liquid Sunlight Alliance, which is funded by the U.S. Department of Energy, Office of Science, Office of Basic Energy Sciences, Fuels from Sunlight Hub under Award Number DE-SC0021266. The authors would like to thank Hong-Zhou Ye for access to the raw data of Ref \citenum{ye2020self} and helpful discussions.

\section*{Conflicts of Interest}
M.H.-G. is a part-owner of Q-Chem, which is the software platform in which the developments described here were implemented.

\bibliography{references}
\end{document}